# Semantic Information Encoding in One Dimensional Time Domain Signals

Kaushik Kumar Majumdar, Systems Science and Informatics Unit, Indian Statistical Institute, 8th Mile, Mysore Road, Bangalore 560059, India; e-mail: kmajumdar@isibang.ac.in.

Srinath Jayachandran, Systems Science and Informatics Unit, Indian Statistical Institute, 8th Mile, Mysore Road, Bangalore 560059, India; e-mail: sri9s@yahoo.in

*Abstract*—A time domain analog signal $s(t)$ is a function $s : \Re^+ \cup \{0\} \to \Re$, where $\Re$ is the set of real numbers and $\Re^+$ is the set of all positive real numbers. For most practical purposes we can assume $s''(t)$, the double derivative of $s(t)$, exists in any closed bounded interval $a \le t \le b$, except at most on a finite number of points. We will show that $s(t)$ can be visualized as the trajectory of a particle moving in a force field with one degree of freedom, in which semantic information is being embedded at any time $t$ at a rate $s''(t)s'(t)$. This holds for the digitized signal $s[n]$ as well, but the derivative becomes difference operation. We have shown here semantic information in the form of the shape of $s[n]$ can be encoded in 13 distinct 3-point configurations, where these 3 points constitute the smallest neighborhood of $n$. It was then shown that semantic information in the form of shape can be encoded in an analog signal in 17 different configurations. A new notion of entropy has also been introduced. We have designed a deterministic finite automaton (DFA) that can accept a finite length digital signal as a regular expression consisting of 13 symbols, i.e., all finite set digital signals form a regular language. We have shown that transducers following from this DFA can detect action potentials and speech phonemes as uttered by different speakers. Finally, we have considered the case $F_s \to \infty$, where $F_s$ is the sample frequency.

## I. INTRODUCTION

Signals are omnipresent in our life, from our brain, heart, mobile phones, radios, televisions to computer networks. Quite sophisticated mathematics is employed in analyzing signals. Yet it is hard to propose a universally acceptable mathematical definition of signal. *Signal* usually means something that conveys information [1], [2], and can be visualized as a function of space (such as, an image) or a function of time (such as, an electrocardiogram) or both [1]. In this work we will remain focused only on time domain signals. Although a precise mathematical definition of a time domain signal is not always implicit, mathematical operations like Fourier and wavelet transforms warrant imposition of specific mathematical conditions on the signal [3], [4], which may eventually become the basis of a mathematical definition. We will see how such a definition can help us understanding the meaning of a signal.

Since a signal is supposed to convey information [1], [2], encoding and decoding of information in the signal are of interest [5], [6]. This information is semantic information [7] as opposed to the Shannon information [1]. Let us take the example of a fair coin toss. It is a random event with probability 0.5 for each of the two outcomes. Its entropy or the average Shannon information content is log 2. But this is before the toss. After the toss there is no randomness or ambiguity. The outcome is already known. It is either head with probability 0 and tail with probability 1 or head with probability 1 and tail with probability 0. So, the Shannon entropy or the average information content of the event is 0 (since $\lim_{x \to 0} x \log x = 0$, we can take $x \log x = 0$ for $x = 0$). However, in this case semantic information is either head or tail and it is nonzero. There is no precise definition for semantic information yet [8]. In this work we go by standard interpretation, i.e., semantic information contained in a signal is its meaning manifested by its shape. In this sense a cardiologist diagnose cardiac arrythmia from the semantic information of the ECG signal or an epileptologist identifies a seizure from the semantic information of the EEG signal.

It has been argued that semantic information is not very useful in practical applications [8]. We have observed in analog signals, the instances in which semantic information encoding is accomplished in the most straightforward manner, do not contain interesting information. Rather, instances of breaking down of the encoding mechanism contain important information. Nevertheless, the current work has shown how much insight can be gathered from the study of semantic information, which may have immense potentiality for applications.

Semantic information encoding is important for signal processing in mobile sensor networks [9], in ECG signals [10], in EEG signals [11] (in both ECG and EEG it is in terms of local shapes), for the study of matching between signals [12], etc. Semantic encoding in images is an essential prerequisite for semantic image understanding, because image understanding is the process of converting "pixels to predicates," i.e., iconic image representations to symbolic form of knowledge [13], [14]. In fact, whenever geometric pattern recognition in a time series is involved, semantic information encoding in the form of shape must be presumed. Visual signal analysis, which for example, is widely practiced on biomedical signals, falls in this category. Semantic information pertains to the actual meaning of the signal as opposed to Shannon theory of information, in which the meaning of the signal is altogether ignored [1], [7].

In this work, borrowing ideas from classical particle mechanics, we will show that if a time domain signal can be visualized as the trajectory of a particle, then information is encoded at each point in the digitized form of that signal as one of 13 different patterns, that gives shape to the smallest neighborhood of that point (in a digital signal smallest neighborhood of a point can have at most three points including the point itself). The signal becomes recognizable by a DFA. Earlier, only ECG signals were shown to be recognizable by automata [10]. We have shown some applications of these 13 patterns in speech and biomedical signal analysis, and have also come up with a new measure of entropy. First, starting with analog signals, going to digitization and then coming back to analog again, we have shown that the analog signals consist of 17 different patterns. In other words, semantic information in the infinitesimal neighborhood of a point in an analog signal can be encoded only in 17 different ways.

The paper is organized in the following way. The next section is devoted to developing basic mathematical notions about a one dimensional time domain signal, all of which are equally applicable to a time series. Section III models discrete signals as strings of a regular language. Section IV deals with semantic information encoding in an analog signal. Section V empirically investigates the relationship of information encoding with sampling rate. Section VI introduces a new notion of entropy based on semantic encoding of information in a digital signal. In the last section we conclude the paper with a discussion on future directions.

## II. Mathematical Preliminaries

### A. Signal as a Function

A one dimensional time domain analog signal is defined on all values of time. Time is nothing but the nonnegative real line. On each time point the signal takes a finite real value. We can express it as $s : \Re^+ \cup \{0\} \to \Re$, where $\Re$ is the set of real numbers and $\Re^+$ is the set of positive real numbers. $\left| s(t) \right| < \infty, \ \forall t \in \Re^+ \cup \{0\}$. Let us make the following assumption about $s(t)$.

*Assumption 1*: For $0 \le a < t < b < \infty$, $s''(t)$ exists at all but at most a finite number of points in $(a, b)$.

We have elaborated in Appendix A why the phrase 'finite number of points' is essential. Assumption 1 implies that $s(t)$ and $s'(t)$ are both continuous at all but at most a finite number of points in $(a, b)$. Also, second order Taylor's expansion of $s(t)$ is possible at all but a finite number of points in $(a, b)$. In other words, at all but at most a finite number of points in $(a, b)$ $s(t)$ can be traversed along a second order polynomial or a parabola. Assumption 1 is also in conformity with the fundamental formula given by equation (1) for reconstruction of the analog signal from its digitization

$$s(t) = \sum_{n=-\infty}^{\infty} s[nT] \frac{\sin(\pi / T)(t - nT)}{(\pi / T)(t - nT)}, \qquad (1)$$

where $T = 1 / F_s$ [3] (p. 388), [15]. The right side of (1) is clearly second order differentiable.

*B. Model*

A one dimensional time domain signal $s(t)$ can freely take the value $s(t)$ for any $t$, that is, it has one degree of freedom. We can model $s(t)$ as the trajectory of a particle moving back and forth in a straight line (it has only one degree of freedom). If the particle has uniform nonzero velocity for all the time, its displacement $s(t)$ will be monotonically increasing or decreasing, leading to violation of the condition $|s(t)| < \infty$ for becoming a signal. Uniform zero velocity will give rise to a constant signal, which does not contain any information, violating the condition laid down in [1] and [2]. So, the particle's momentum must be changing with respect to time in general. By Newton's second law of motion it must be moving under the influence of a force, which is given by $s''(t)$, assuming the mass of the particle to be unit.

Now, if the particle makes an infinitesimal displacement $ds$ the work done by it is $s''(t)ds(t)$. Let the execution of this infinitesimal work takes infinitesimal time $dt$ to be accomplished. Therefore, the rate at which the work is done is the power

$$P(s(t)) = s''(t)s'(t), \qquad (2)$$

where P denotes the *power-operator* or *P-operator*. It is clear from Assumption 1 that in any open interval $(a,b)$ P-operation is valid at all but at most a finite number of points in the interval. In other words, given any open interval $(a,b)$ we always have a finite decomposition of $(a,b)$ into $k$ number of open subintervals, such that, the following holds,

$$(a,b) = \bigcup_{i=1}^{k} (a_i, b_i), \qquad (3)$$

where $(a_i, b_i) \bigcap (a_j, b_j) = \phi$ for $i \neq j$, $\phi$ is the empty set, and $s''(t)$ may not exist for $t = a_i$ or $t = b_i$, $i \in \{1, ...., k\}$.

It is obvious from equation (3) that given a signal $s(t)$ there exist open intervals $(a', b') \subset \Re^+ \cup \{0\}$, such that, for $t \in (a', b')$ $s''(t)$ exists. The 'force field' $s''(t)$ gives rise to (generates) the signal $s(t)$, in the sense that, if $s''(t)$ is applied on a particle, which is capable to move only along a straight line under the influence of $s''(t)$, then the particle will always make $s(t)$ displacement along the line in time $t$. We will get the signal on a paper or on a tape, like paper EEG or paper ECG or speech signal on an oscilloscope, if we plot the graph $(t, s(t))$. In other words, if $s''(t)$ exists for $t \in (a', b')$, the one dimensional time domain signal $s(t)$ can always be modeled as the trajectory of a particle moving in a force field with one degree of freedom within the interval $(a', b')$.

By (3) it is clear that if Assumption 1 holds, a one dimensional time domain signal can be modeled as the countable union of piecewise trajectories of a moving particle in a force field with one degree of freedom.

*C. Semantic Information*

In this paper we are concerned about the meaning of a signal as revealed by its geometric shape as opposed to the information content by entropy measure (we will of course discuss about entropy later). Semantic information from a signal can be retrieved in many different ways including Fourier and wavelet transformations [4], [18]. Here we introduce a way of encoding information at a point in a one dimensional time domain signal, when the temporal double derivative exists at that point.

Equation (2) gives the rate at which work is being done by the moving particle in a force field. This is also the rate at which the kinetic energy of the particle is being dissipated to give shape to the trajectory $s(t)$ of the particle. In other words, the kinetic energy of the particle is being dissipated to create the unique geometric pattern of the signal $s(t)$, which holds the meaning or the *semantic information* of the signal in terms of its shape in the infinitesimal neighborhood of $t$. Kinematics of a particle and evolution of a one dimensional time domain signal are therefore similar to each other, which holds the key to embedding semantic information into the signal. Therefore, equation (2) gives the rate at which semantic information is encoded at the point $t$. The total semantic information content $I(s(t), a, b)$ of the signal $s(t)$ in an interval $(a, b)$ is then given by

$$I(s(t), a, b) = \int_a^b \left| P(s(t)) \right| dt . \qquad (4)$$

Here we are not distinguishing between noise and signal, as noise is a relative term, depending on the specific application. Whatever time series we are getting, we are treating that as a pure signal.

*D. Digital Signal*

Assuming we have digitized the analog signal $s(t)$ at or above the Nyquist rate to get the digital signal $s[n]$. How to encode the semantic information in terms of the geometric shape of $s[n]$ in the smallest neighborhood of $n$, i.e., in the set $\{n-1, n, n+1\}$, where $n$ is an interior and not a boundary point?

Equation (4) gives one type of quantification of semantic information in the signal segment of $s(t)$ between $a$ and $b$. However, this is not very useful. Also P-operator works as a high-pass filter and suppresses components associated with less than 1 Hz (can be easily verified for sinusoidal signals). Instead, if we study how $P(s(t))$ is changing sign or retaining the same sign at point $t$, we get precise information about the shape of the signal $s(t)$ in an infinitesimal neighborhood of $t$. It is particularly instructive for a digital signal $s[n]$ and a discrete version of P-operator $P(s[n])$.

*Theorem 1*: $s[n]$ can be in one of 13 different geometric configurations as listed in Table I in the smallest neighborhood of $n$, in which $n$ is an interior point.

Table I

List of all smallest neighborhood information encoding configurations in a digital signal

| Sign change | s'[n] | s''[n] | s'[n+1] | 3-point configuration | Configuration number | Empirical observation |
|---|---|---|---|---|---|---|
| -- | - | + | - | | 1 | |
|  | + | - | + | | 2 | |
| ++ | + | + | + | | 3 | Statistically more abundant |
|  | - | - | - | | 4 | |
| -+ | - | + | + | | 5 | |
|  | + | - | - | | 6 | |
| 00 | + | 0 | + | | 7 | Statistically less abundant |
|  | - | 0 | - | | 8 | |
|  | 0 | 0 | 0 | | 9 | |
| 0+ | 0 | + | + | | 10 | |
|  | 0 | - | - | | 11 | |
| -0 | + | - | 0 | | 12 | |
|  | - | + | 0 | | 13 | |

Studying different datasets we have observed that some of the configurations are relatively more abundant and some are less (see Fig. 1 for an example).

Before proving Theorem 1 we will establish a hierarchy of sign order for the P-operator. By the hierarchy of sign order we mean the following sign sequence in increasing order.

$$- \quad < \quad 0 \quad < \quad + \tag{5}$$

P-operator never changes sign from higher order to lower order, e.g., P-operator never changes sign from positive to negative as shown in Lemma 1. Lemma 2 and Lemma 3 show that P-operator cannot change sign from zero to negative or from positive to zero respectively. In subsection II(B) we have seen that the kinetic energy of a moving particle is transformed into semantic information at each point of the trajectory or the signal, in the form of its shape in an infinitesimal neighborhood of that point. Lemma 1, Lemma 2 and Lemma 3 together assert that once information is encoded, it never transforms back to the kinetic energy, which intuitively make sense. This also justifies considering sign changes of P-operator from the left product to the right product (Fig. 2), instead of the numerical value, for identifying semantic information encoded into the signal.

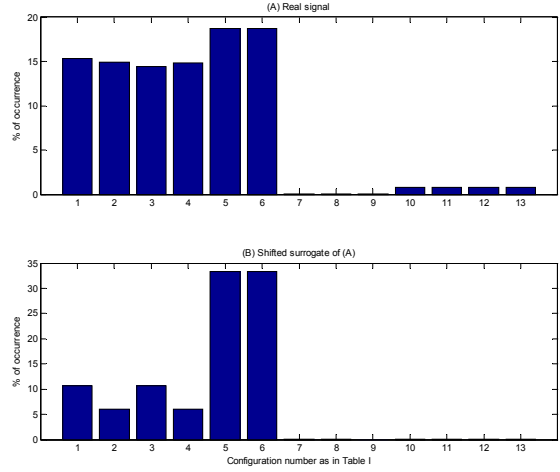

Fig. 1. Histogram plot of % of occurrence of 13 configurations as listed in Table I vs. configuration number for (A) real signal (top) and (B) shifted surrogate of the real signal (bottom). Configurations numbered 7 through 13 are relatively rare in both the signals as noted in Table I. Signals are one hour long with sample frequency 256 Hz.

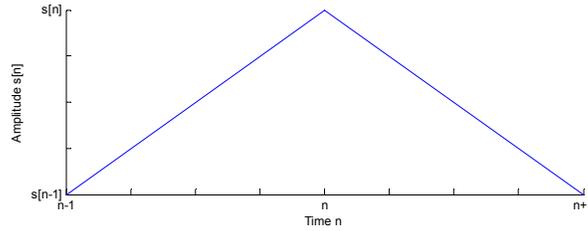

Fig. 2. Time $n$ vs. amplitude $s[n]$ at $n$. *Left product* of $P(s[n])$ is $P(s[n-]) = s''[n] * s'[n]$ and *right product* of $P(s[n])$ is $P(s[n+]) = s''[n] * s'[n+1]$, where $s'[n] = s[n] - s[n-1]$ and $s''[n] = s'[n+1] - s'[n]$.

Fig 2 clearly shows that the first difference is backward difference and the second difference is forward difference. The two conventions are not mutually contradictory. We must follow this to get the value of P-operator at the point $n$ in the neighborhood $n-1$, $n$, $n+1$. If we follow the same convention for both the difference operations, we will get the value of P-operator either at $n-1$ or at $n+1$, but not at $n$.

*Lemma 1*: $P(s[n-]) > 0$ and $P(s[n+]) < 0$ is impossible to occur.

*Proof*: Let us consider the smallest neighborhood of $n$ consisting of only $n-1$, $n$ and $n+1$, for which $P(s[n-]) > 0$ and $P(s[n+]) < 0$ hold. Since the P-operator is changing sign from positive left product to negative right product (see Fig. 2 for the products) the following two cases are possible.

Case 1: $s''[n] > 0$ and $s'[n] > 0$, and $s''[n] > 0$ and $s'[n] < 0$. Since $s'[n] > 0$ and $s'[n+1] < 0$, $s[n] > s[n-1]$ and

$s[n+1] < s[n]$, which implies $2s[n] > s[n-1] + s[n+1]$. But $s''[n] = s[n+1] - 2s[n] + s[n-1] > 0$, which implies

$s[n+1] + s[n-1] > 2s[n]$, leading to a contradiction.

Case 2: $s''[n] < 0$ and $s'[n] < 0$, and $s''[n] < 0$ and $s'[n+1] > 0$. Same way as in Case 1, this too will lead to a contradiction. This completes the proof.

*Lemma 2*: $P(s[n-]) = 0$ and $P(s[n+]) < 0$ is impossible to occur.

*Proof*: $s''[n] \neq 0$, because $s''[n]$ is a factor common in both $P(s[n-])$ and $P(s[n+])$, and $P(s[n+]) \neq 0$. The following two cases are possible.

Case 1: $s'[n] = 0$, $s'[n+1] < 0$ and $s''[n] > 0$.

$$s'[n] = 0 \Rightarrow s[n-1] = s[n]. \qquad (6)$$

$$s'[n+1] < 0 \Rightarrow s[n+1] < s[n]. \qquad (7)$$

$$s''[n] = s[n+1] - 2s[n] + s[n-1] > 0$$

$$\Rightarrow s[n+1] + s[n-1] > 2s[n]. \qquad (8)$$

(6) + (7) gives $s[n+1] + s[n-1] < 2s[n]$, which is contradictory to (8). So, Case 1 is impossible.

Case 2: $s'[n] = 0$, $s'[n+1] > 0$ and $s''[n] < 0$.

In a similar way as in Case 1 it can be shown that Case 2 is also impossible. This completes the proof of Lemma 2.

*Lemma 3*: $P(s[n-]) > 0$ and $P(s[n+]) = 0$ is impossible to occur.

*Proof*: Same as in Lemma 2.

*Proof of Theorem 1*: Table 1 gives us the list of sign changes or preserving the same sign from left product to right product in P-

operator. Both the products are product of two factors. Also, the same double difference factor is common in both of them.

Therefore, there are three distinct factors making up all two of them. Each of these factors can take one of the three signs shown in (5). So, there are at most $3^3 = 27$ possibilities.

Lemma 1, Lemma 2 and Lemma 3 rule out two possibilities each leading to reduction of 6 possibilities out of 27. We are now left with only 21 possibilities.

Left product of P-operator is 0 and right product of P-operator is also 0 can happen in 11 different ways, but only the following three case are possible to occur.

1. $s'[n] > 0, \quad s''[n] = 0, \quad s'[n+1] > 0$.

2. $s'[n] < 0, \quad s''[n] = 0, \quad s'[n+1] < 0$.

3. $s'[n] = 0, \quad s''[n] = 0, \quad s'[n+1] = 0$.

It is easy to check 1, 2, 3 are valid occurrences. Now, if $s''[n] \neq 0$, both $s'[n]$ and $s'[n+1]$ must vanish. This can happen in two different ways and both can be ruled out, because the following two cases are impossible.

Case 1: $s''[n] > 0, \quad s'[n] = 0, \quad s'[n+1] = 0$.

Case 2: $s''[n] < 0, \quad s'[n] = 0, \quad s'[n+1] = 0$.

$s''[n] > 0 \Rightarrow 2s[n] > s[n-1] + s[n+1]$, and $s'[n] = 0$, $s'[n+1] = 0$ together imply $2s[n] = s[n-1] + s[n+1]$, leading to a contradiction. Therefore, Case 1 is not possible to occur. Similarly, it can be shown Case 2 is also impossible to occur. It is easy to verify that the following six cases are not possible either.

(a) $s''[n] = 0, \quad s'[n] < 0, \quad s'[n+1] > 0$.

(b) $s''[n] = 0, \quad s'[n] > 0, \quad s'[n+1] < 0$.

(c) $s'[n] > 0, \quad s''[n] = 0, \quad s'[n+1] = 0$.

(d) $s'[n] < 0, \quad s''[n] = 0, \quad s'[n+1] = 0$.

(e) $s'[n] = 0, \quad s''[n] = 0, \quad s'[n+1] > 0$.

(f) $s'[n] = 0, \quad s''[n] = 0, \quad s'[n+1] < 0$.

With this we have shown that 8 more cases out of 21 are not possible to occur. Therefore, there are only 13 cases, all of which are listed according to valid change of sign from left product to right product in the factorization of P-operator (all of them actually occur in a digital has been shown in Fig. 1). This completes the proof of Theorem 1.

*Corollary 1*: A peak or a trough is a digital signal is the most information rich event.

*Proof*: Notice that in Table 1 a peak or a trough occurs in a digital signal only when P-operator changes sign from negative to positive (this is true of an analog signal as well). It will be a peak if $s''[n] < 0$ and a trough when $s''[n] > 0$. According to the sign hierarchy in (5) changing from negative to positive requires two jumps, which only happen when there is a peak or a trough in the signal. Any of the other 11 features requires only one jump in (5) as can be seen in Table 1. Therefore, a peak or a trough is the most information rich event in a signal. Notice that this proof can be extended in a straight forward manner to an analog signal too.

It is clear that both the proofs of Theorem 1 and Corollary 1 can be extended to multirate signals in a straight forward manner. The assertion of Corollary 1 can also be verified from Fig. 1. In a randomized signal (Fig. 1(B)) % of peaks (configuration 6 in Table 1) and % of troughs (configuration 5 in Table 1) are higher than the nonrandom signals (Fig. 1(A)). It is well known that a randomized signal's entropy or information content is higher than a nonrandom signal. It is clear form Table 1 and Fig. 1 that configurations 7 through 13 are relatively rare in signals, For each of them P-operator becomes zero either during left product or during right product.

### III. Signals as Language

#### A. Deterministic Finite Automaton

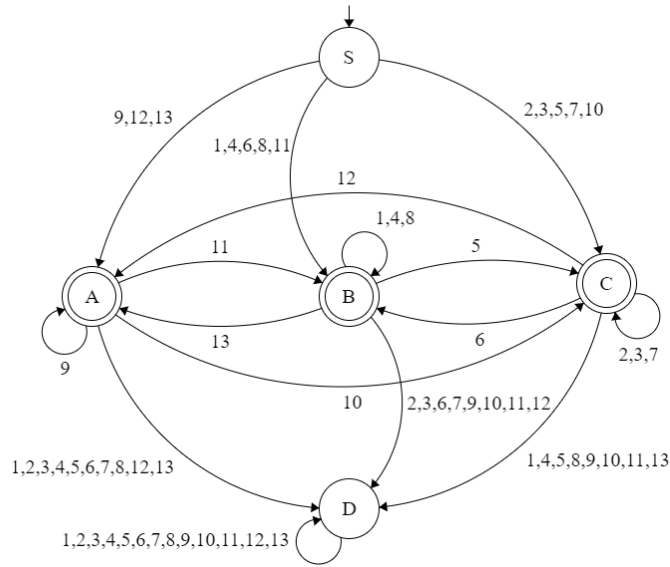

Fig. 3. Transition diagram of the DFA that recognizes any digital signal of finite length as a valid string of 13 symbols or configurations as given and numbered in Table 1. Accepting states A, B and C have been elaborated in Table 2. This is the DFA that recognizes the collection of all finite length digital signals as a regular expression (see ref. [16]).

In the previous section we have shown that a digital signal can be represented as a string of 13 configurations or characters.

From this it is quite tempting to represent digital signals as members of a language (see ref. [16] for a formal definition of a language). In fact, digital signals have already been represented as language, but in a different way. Here we will take a more direct approach to represent a digital signal as a regular expression recognizable by a deterministic finite automaton (DFA) as shown in Fig. 3. The DFA in Fig. 3 is defined by the quintuplet

$$M(s) = \left(Q, \Sigma, \delta, S, F\right),\qquad\qquad(9)$$

where $M(s)$ is the DFA that accepts any finite length digital signal $s$ (all real signals are of finite length),

$$Q = \{S, A, B, C, D\}\qquad\qquad(10)$$

is the set of states (accepting states $A$, $B$ and $C$ have been elaborated in Table 2), $\Sigma$ is the alphabet of 13 configurations given in Table 1 (we will denote them by the number assigned in Table 1), $\delta$ is the transition function given by the transition Table 3, $S$ is the start state and $F = \{A, B, C\}$ is the set of accepting states $A$, $B$ and $C$ as shown in Fig. 3. A dead state $D$ has been added to signify invalid transitions.

Table 2

Accepting states $A$, $B$ and $C$ and the next transition

| $A$ Configurations ending with 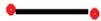 | Configurations starting with 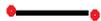 |
|---|---|
| <u>9</u>, 12, 13 | <u>9</u>, 11, 10 |
| $B$ Configurations ending with 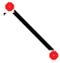 | Configurations starting with 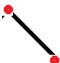 |
| <u>1</u>, <u>4</u>, 6, <u>8</u>, 11 | <u>1</u>, <u>4</u>, 5, <u>8</u>, 13 |
| $C$ Configurations ending with 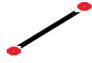 | Configurations starting with 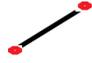 |
| <u>2</u>, <u>3</u>, 5, <u>7</u>, 10 | <u>2</u>, <u>3</u>, 6, <u>7</u>, 12 |

For a valid transition, preceding configuration's last part (line joining the last two points) and the following configuration's first part (line joining the first two points) must be the same.

Table 3

The transition table of the DFA in Fig. 3

|      | 1 | 2 | 3 | 4 | 5 | 6 | 7 | 8 | 9 | 10 | 11 | 12 | 13 |
|------|---|---|---|---|---|---|---|---|---|----|----|----|----|
| →S   | B | C | C | B | C | B | C | B | A | C  | B  | A  | A  |
| *A   | D | D | D | D | D | D | D | D | A | C  | B  | D  | D  |
| *B   | B | D | D | B | C | D | D | B | D | D  | D  | D  | A  |
| *C   | D | C | C | D | D | B | C | D | D | D  | D  | A  | D  |
| D    | D | D | D | D | D | D | D | D | D | D  | D  | D  | D  |

All transitions ending in an accepting state have been shaded. * on top left denotes accepting state. Arrow indicates the start state.

Strings of symbols that are accepted by a DFA are called *regular expressions* [16]. Collection of all regular expressions accepted by a DFA is called a *regular language* [16]. We have already established that the collection of all finite length digital signals $s$ constitute a regular language, because all those signals are accepted by the DFA $M(s)$ given by (9). $M(s)$ is a very general DFA accepting all real life digital signals. For a specific class of signals it is possible to come up with a more specialized DFA as in case of neuronal action potentials given by the Hodgkin-Huxley or HH model [19] (Chapter 6). The rare features 7 through 13 do not appear in the simulation of HH model. Let $H(a)$ be the DFA that can accept action potential signals $a$. However, action potentials are fixed amplitude impulses. The $H(a)$ can only capture the shape of an action potential, but not its size. Therefore we need a more specialized recognizer for an action potential, which will generalize $H(a)$ by incorporating the notion of amplitude into it.

*B.  Transducer*

We actually need a DFA with an output or a *transducer* for recognizing the pattern of an action potential. An action potential simulated by HH model and the transition diagram of the transducer $H_t(a)$ have been shown in Fig. 4(A) and Fig. 4(B) respectively.

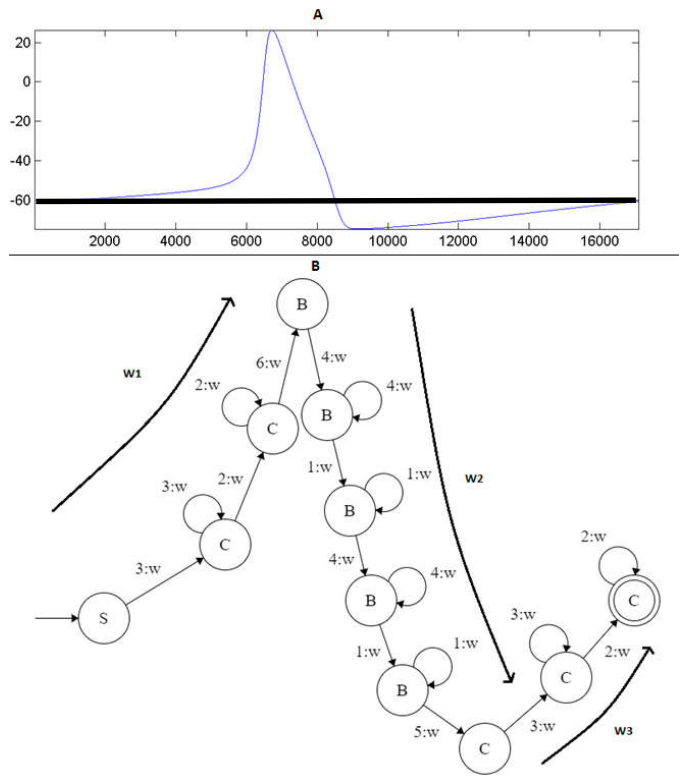

Fig. 4. (A) An action potential $a$ simulated by HH model. Thick horizontal line indicates the threshold potential. (B) The transition diagram of the transducer $H_t(a)$ for recognizing the action potential $a$. The output weight $w$ takes a large, but finite, number of values as shown in Fig. 5. **w1** is the cumulative weight during depolarization, **w2** is the cumulative weight during hyperpolarization and after-hyperpolarization, and **w3** is the cumulative weight during the undershot.

Transducers, or rather a specialized form of it, called *weighted finite state transducers* (WFSTs) have been used profitably for recognizing speech signals [20], [21]. Here we have shown with an example how WFSTs can be used to recognize particular events or occurrences in a digital signal. $H_t(a)$ in Fig. 4(B) is a WFST. It has only 6 input symbols (1 through 6) as shown in Fig. 4(B) and Fig. 5. Accepting states A, B and C are as shown in Table 2, i.e., the line segment joining the last two points is either having zero slope (state A) or negative slope (state B) or positive slope (state C). The weight associated with the transition to the next state is $w$, where $w = \{\tan^{-1}(b/a)\}/90$. Clearly, $w \in [0,1)$. As elaborated in Fig. 5, the weight is a normalized measure of the increase (or decrease) in amplitude $b$ of the digital signal from one time point to the next, i.e., $b = s[n+1] - s[n]$, where $w = (\tan^{-1} b)/90$, as $a$ is taken to be 1 (in multirate signal the actual value of $a$ will have to be taken).

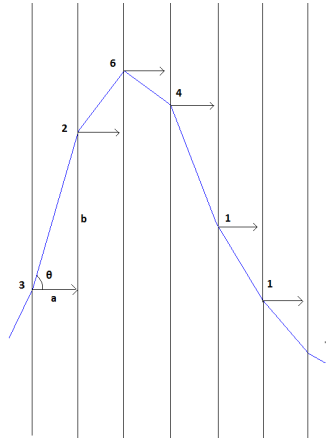

Fig. 5. Scheme for fixing the value of weight $w$. $w = \theta / 90 \in [0,1)$, where $\theta = \tan^{-1}(b/a)$. The digitized signal $s[n]$ is represented by the piecewise blue lines, where the number represents the configuration in Table 1. $b = s[n+1] - s[n]$ and $a = 1$.

In Fig. 4(B), when the signal segment is an action potential the relation

$$\mathbf{w2} - \mathbf{w1} \approx \mathbf{w3}, \qquad\qquad\qquad (11)$$

holds. $\mathbf{w1}$, $\mathbf{w2}$ and $\mathbf{w3}$ have been defined in the caption of Fig. 4. This can easily be proved. When the sample frequency is high (for the spike train signals it is higher than 10 kHz), $\lim_{n \to \infty} s[n+1] - s[n] = b \to 0$, which implies $\theta = \tan^{-1}(b/a) = \tan^{-1} b = b$ (since $a = 1$), i.e., $w = b/90$. From this and Fig. 5 it is clear that the vertical distance of the highest point of action potential above the threshold line as shown in Fig. 4(A) and is equal to $90 * \mathbf{w1}$ (see Fig. 4(B) for $\mathbf{w1}$). The vertical distance from the highest to the lowest point of the action potential is $90 * \mathbf{w2}$ and the vertical distance of the lowest point of the action potential from the threshold line is $90 * \mathbf{w3}$. So, $90 * \mathbf{w2} - 90 * \mathbf{w1} = 90 * \mathbf{w3}$ or $\mathbf{w2} - \mathbf{w1} = \mathbf{w3}$. In general, the undershot of an action potential may not reach the threshold in order to initiate the next action potential (the so called pacemaker effect). Therefore, $\mathbf{w2} - \mathbf{w1} = \mathbf{w3}$ is replaced by $\mathbf{w2} - \mathbf{w1} \approx \mathbf{w3}$. This proves that the condition (11) is necessary for the signal segment to be an action potential.

However, this condition is not sufficient as can be seen by taking a signal segment whose shape and amplitude (both below and above the threshold line) are equal to an action potential but whose duration is much larger. This can be rectified by taking a transducer which maps a string of symbols to a string of weights, i.e., a weight vector. If $w_n$ is the $n$ th entry of the weight vector, then $w_n$ is the output weight at the $n$ th transition. In short, $S_n a_n : w_n$, where $S_n$ is the state at the beginning of the $n$ th transition and $a_n$ is the input symbol at the $n$ th transition, i.e., the $n$ th entry of the input string. (11) will hold in a straight forward manner.

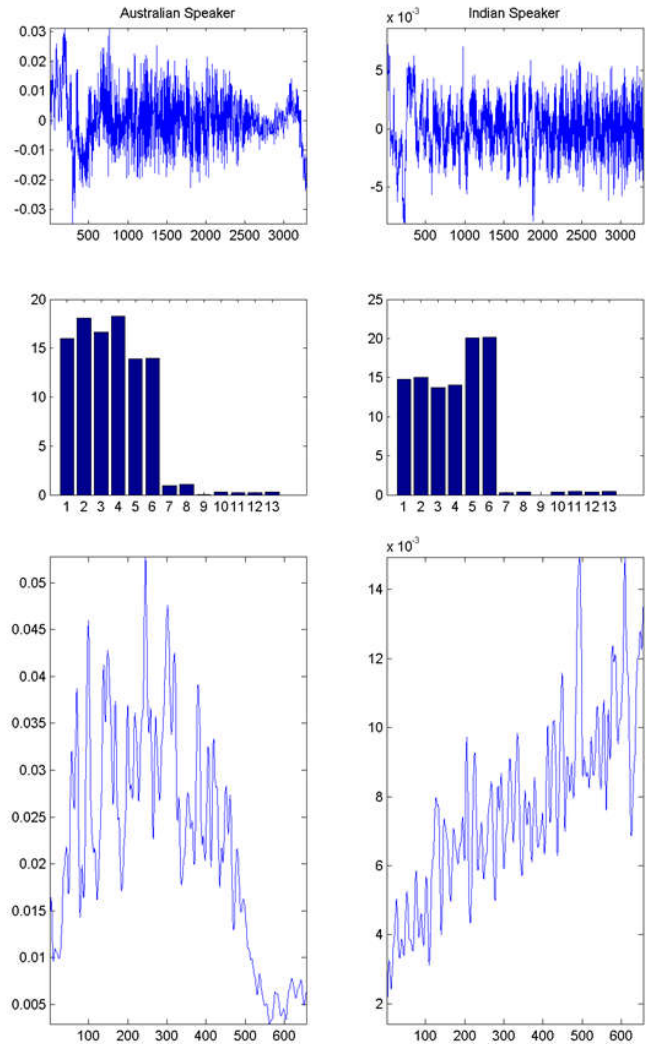

Fig. 6. (Top) The speech signal produced by an Australian male speaker (left) and that of an Indian male speaker (right) during utterance of the phoneme /f/ in the word 'coffee'. (Middle) Histogram of percentage distribution of the 13 configurations in the speech signal uttered by the Australian speaker (left) and that of the Indian speaker (right). (Bottom) Weighted output of the transducer used to differentiate the Australian speaker (left) from the Indian speaker (right) when both are uttering /f/ in 'coffee'.

Fig. 6 demonstrates an application of WFST in recognizing the phoneme /f/ in the word 'coffee' by an Australian male speaker (for whom English is the mother tongue) and an Indian male speaker (for whom English is the second language). Phonetic script for the word "coffee": /'k ɒ f / as extracted by the software package Praat [22]. We have calculated Bhattacharya distance [23] between the frequency distributions of the 13 features during utterance of each of the four phonemes by the Australian speaker and the Indian speaker. The difference between the two speakers is the lowest during uttering the phoneme /f/ in 'coffee'. Yet, the transducer outputs are quite different as shown in the bottom panel of Fig. 6. The transition diagram of the corresponding DFA will be the same as the transition diagram of the DFA for the digital signals as shown in Fig. 3. However, the transducer for the phoneme signals will have different vectors of output weights for different speakers.

## IV. ANALOG SIGNAL

It appears that the distribution of 13 features of Table 1 in any digital signal segment will depend on the sample frequency. However, Assumption 1 implies that in any finite signal segment there are only a finite number of points where the analog signal is double differentiable. In between any two such successive points $a_1$ and $b_1$ ( $a_1 < b_1$) the double differential of the signal can be negative, positive, zero. The segment on which the double derivative is positive the signal is convex. The segment on which the double derivative is negative the signal is concave and the segment on which the double derivative is zero the signal is a straight line (Fig. 7). In $(a_1, b_1)$ the signal can have either of the seven fundamental shapes, elaborated in Fig. 7, or a combination of those shapes in which the boundary point between any two adjoining shapes also admits double derivative. In an analog signal these shapes are *fundamental* in the sense that decomposing any of these shapes gives back components with that shape only. In other words, increasing sample frequency will only increase the number of discrete 3-point configurations used to approximate each of those seven shapes and will not bring any change in the type of configurations. In Fig. 7 it has been elaborated that each of the seven shapes is represented by only one of the thirteen configurations in Table 1, when sampled more than two times.

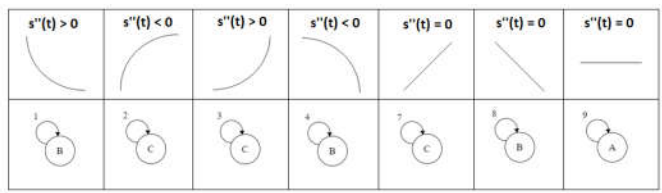

Fig. 7. (Top) Seven fundamental shapes appearing in any analog signal where the signal is double differentiable. (Bottom) The corresponding 3-point configuration (expressed by its number in Table 1) by which the fundamental shape right above can be approximated when sampled more than two times and the transition diagram of the approximation of the shape by the 3-point configuration right below for any number of sampling greater than three.

It is clear that at any point of the analog signal interior to any of the seven shapes the signal is double differentiable. Let us call all such points as *doubly smooth point* or for simplicity, *smooth point* only. All points where the signal is not differentiable must occur at the boundary of the seven shapes (even boundary points can some time be smooth). Let us call all such points as *break points*. An analog signal consists of (doubly) smooth points and break points only. In Appendix A we have shown the break points cannot have a limit point under Assumption 1 and therefore they must be isolated.

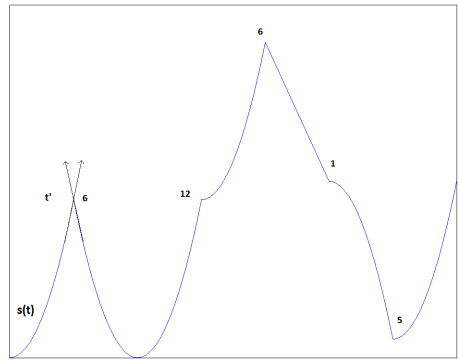

Fig. 8. An analog signal $s(t)$, in which the shape or configuration of an infinitesimal neighborhood of the break points has been indicated by the corresponding number in Table 1. Tangent to the left cusp and to the right cusp having boundary at the break point $t'$ have been shown by arrow.

When a break point appears in the boundary in between two shapes in Fig. 7, that break point becomes a point of intersection of the two tangents respectively to the two shapes at the break point (Fig. 8). Now, let us sample the analog signal and let the sample frequency $F_s \to \infty$. If the break point is at $t'$, so that $s'(t')$ does not exist, then the two tangents intersecting at $t'$ will have infinitesimally vanishing length under the sampling and will constitute the smallest 3-point neighborhood of $s(t')$ at $t'$. That neighborhood can take only 10 out of the 13 configurations in Table 1. Only configurations 7, 8 and 9 are not possible, because $t'$ will then become an interior point to one of the three right most shapes in Fig. 7, leading to a contradiction to $t'$ is a boundary point. This proves in an infinitesimal neighborhood of a break point $t'$ an analog signal $s(t)$ can have only 10 shapes or configurations listed in Table 1 except configurations 7, 8 and 9. In other words, $s(t)$ is composed of 7 shapes in Fig. 7 and 10 shapes in Table 1. By now we have proved the following,

*Theorem 2*: Semantic information can be encoded in a one dimensional time domain analog signal in terms of 17 shapes described in Fig. 7 and Table 1 (except configurations 7, 8 and 9, analog of which are appearing in Fig. 7).

It is clear, that in order to preserve information in an analog signal during digitization, sparse sampling of the shapes in Fig. 7 and dense sampling of the shapes in Table 1 will have to be followed. By *dense sampling* we mean leaving out no samples. If we allow $F_s \to \infty$, it is clear from the transition diagrams presented in the bottom panel of Fig. 7, that the shapes in Fig. 7 will all be replaced by the shapes in Table 1.

## V. SAMPLING RATE

Up sampling and down sampling of a digital signal can be performed in multiple ways. The scheme that we followed here is to reconstruct the analog signal out of the digital by cubic spline fitting [24] and resampling. We have applied it to the simulated action potential by HH model (Fig. 9). We started with 4000 kHz sample frequency and came down to 2 kHz sample frequency. In a spike train Nyquist rate cannot exceed 2 kHz, because action potentials are never less than 1 ms wide. It is evident from Fig. 9 that distribution of the 13 configuration didn't change appreciably throughout the variation of $F_s$ from 4000 kHz to 2 kHz. In case of an electrode artifact in human depth EEG recording acquisitioned at 256 Hz sampling rate the variation in the distribution of the 13 configurations has been shown in Fig. 10. The percentage of peaks and troughs (configurations 5 and 6 respectively) has started increasing with decreasing sample frequencies. Compared to the HH action potential signal this signal is lot more jittery and therefore has greater percentage of peaks and troughs than the former. However, fitting (cubic) spline on the initial digital signal (sample frequency 256 Hz) makes it less jittery or less noisy [25]. Since after spline fitting the signal becomes smooth, it mostly contains the left most four cusp shapes in Fig. 7, and at high sample frequency produces configurations 1, 2, 3 and 4 of Table 1 as clearly shown in the bottom panel of Fig. 7. This is evident in both Fig. 9 and Fig. 10. Because of smoothing of jitters in the depth EEG signal through spline fitting, much larger number of tiny cusps were created in this signal than the purely generated action potential signal as can be seen comparing Fig. 10 with Fig. 9. As sample frequency went down the number of configurations 1, 2, 3, and 4 inside the cusps went down, but number of peaks (configuration 6) and number of troughs (configuration 5) inside them remained fixed at one at the most in each cusp. This is the reason percentage distributions of peaks and troughs appear to have increased with decreased sample frequency in Fig. 10.

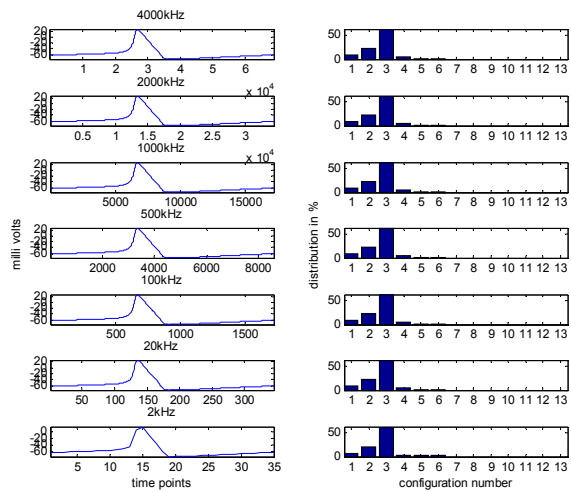

Fig. 9. On the left shown simulated HH action potential and its sampling at rate varying from 4000 kHz down to 2 kHz. On the right shown the histogram of percentage distribution of the 13 configurations in Table 1 in the HH action potential for each sample frequency.

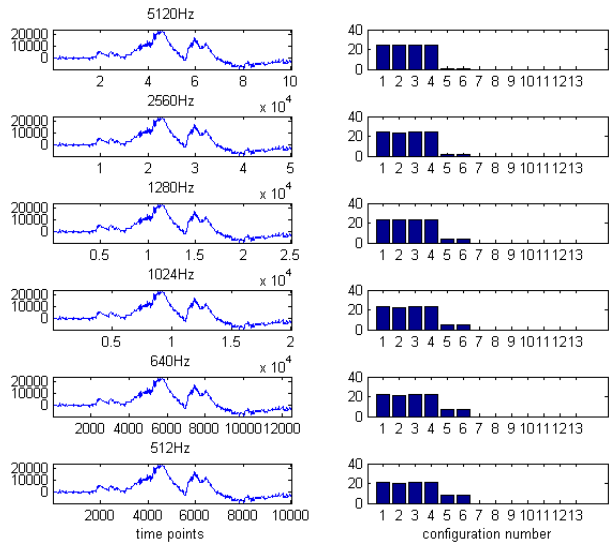

Fig. 10. Same as in Fig 9, but for an electrode artifact in human depth EEG signal. The original sample frequency was 256 Hz.

## VI. Entropy Estimation

Estimating entropy of a real signal is a computationally tricky job [26]. For a time series histogram method is the easiest, but subjective. The resultant entropy value depends on the bin size. Taking clue from permutation entropy [27] we can define an entropy measure for digital signals according to the frequency distribution of the 13 configurations in Table 1. Let us call this measure as *semantic entropy* (SE). Formally,

$$SE(s) = -\sum_{i=1}^{13} p(i) \log_2 p(i), \qquad (12)$$

where $s$ is the signal segment, $p(i)$ is the frequency distribution (actually density) of the $i$ th configuration in Table 1.

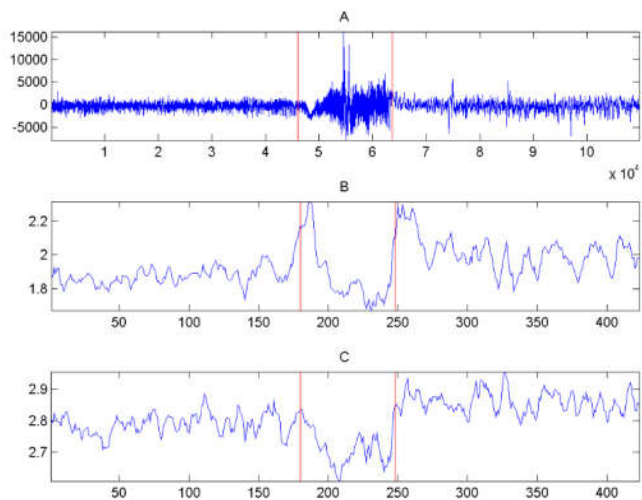

Fig. 11. (A) Raw focal channel EEG signal (time vs. amplitude plot) from a patient with epilepsy during a seizure. (B) Permutation entropy of order 3 of the signal. (C) Semantic entropy of the same. Vertical lines indicate seizure onset and offset points in each plot.

In Fig. 11 we have summarized our brief comparative study between permutation entropy and semantic entropy. Shannon entropy didn't show any trend and we haven't presented the result. Here we have presented results on permutation entropy of order 3, but the same for order 10 and 13 showed identical trend. The trends between the permutation and the semantic entropy are quite similar, except in the beginning of the seizure, when the permutation entropy becomes high. However, a literature search reveals that entropy actually goes down during an epileptic seizure [28].

## VII. CONCLUSION

In this work we have visualized a one dimensional time domain analog signal as the trajectory of a moving particle in a force field with one degree of freedom. Since there is a force field responsible for the motion of the particle, which is engendering the signal, second differentiability of the signal is of predominant importance. We have shown sign of first and second difference operations leads to determining the shape of the smallest neighborhood of a point in a discrete signal. We have mathematically proved that such neighborhoods contain only 3 points and can have 13 distinct shapes. Subsequently we have proved that any infinitesimal neighborhood of a point in an analog signal can take only 17 distinct shapes. A sequential arrangement of these shapes constitutes the whole signal that bears its meaning or the semantic information content. We have also defined semantic entropy. Our next goal is to extend this work to two dimensional signals, i.e., images.

It is interesting to note that on a smooth point of an analog signal P-operator is applicable in a straightforward manner, but the geometric shape or the encoded semantic information is a monotonous cusp or a straight line (Fig. 7). Break points on the other hand cannot be predicted from their neighborhood with the help of deterministic mathematics. Their occurrence is random and isolated. They contain novel information and can be studied statistically. Although information encoding in them can be modeled by discrete P-operator, ambiguity about their temporal evolution can be modeled only by probability distribution. This is one major limitation of the semantic information of a time series. The observations made in [8] seem to be true, which is further supported by the fact that equation (4) is not really very useful. However, we have demonstrated in this work the usefulness of studying semantic information for practically meaningful purposes.

Since digital signals can be expressed as a string of 13 distinct configurations, we introduced a deterministic finite automaton, which can accept any digital signal as a regular expression. We subsequently introduced a weighted transducer whose output weights give an estimate of the amplitude of the signal (13 configurations only give information about the shape of the signal). We have shown applications of this transducer in speech and neuronal signal processing. Fine tuning of this transducer is likely to give very efficient speaker identification and neuronal spike sorting algorithms.

Whatever conceptual and theoretical developments we have accomplished here are applicable to any time series. We have shown that an analog time series contains only two types of points namely, smooth point and break point. Smooth point is one in whose neighborhood the signal is differentiable and break point is one at which derivative does not exist. These two types of points should be sampled differently. In a time series a break point with a high value of double difference operation (such as $t'$ in Fig. 8) may indicate a rare event [29], [30]. More precisely, probability of having a value of double difference at a break point which is greater than the expected value of double differences at break points in a time series is likely to be an important indicator of an unusual behavior. Conditional probability of a large value of double difference at a break point given such large values have already occurred will be appropriate for modeling abnormal spiking activities in earth quake and epileptic seizure signals.

Traceability of a signal is an important property which is often overlooked. Traceability is equivalent to drawing (the graph $(t, s(t))$ of) the signal. This means functions like everywhere continuous but nowhere differentiable cannot qualify as signals. Also signal visualization is important in many applications. The signals which cannot be drawn or plotted cannot be visualized. We have shown in Appendix A that it is essential for $s(t)$ to be second differentiable in order to be viewable. In other words, local second order polynomial approximation at each point of the signal, with at most a finite number of exceptions, must hold. Third order polynomial approximation is taken to be an optimally efficient way of signal reconstruction from its sampled values [24], [31]. It would be an interesting study to explore the relationship between second order approximability and third order reconstruction of the signals.

APPENDIX A

An analog signal should be a continuously traceable function, else its graph cannot be drawn on the plane and clearly it cannot be modeled as the trajectory of a moving particle. Without this property signals cannot be displayed in oscilloscope or traced on a paper like paper-EEG or paper-ECG.

*Definition A1*: An analog signal $s(t)$ is *continuously traceable* at $t$, if and only if in an infinitesimal neighborhood of $t$ the tangent to $s(t)$ and secant of $s(t)$ coincide. In other words, the points $(t, s(t))$ and $(t + h, s(t + h))$, $h \to 0$ on the plane must both be on the graph of $s(t)$.

In Definition A1 $h$ can take both negative and positive signs. In special cases where $h$ can take only negative or positive sign, we call $s(t)$ is (continuously) traceable at $t$ from left or from right respectively. Points that are traceable from the left and from the right with different values for $\lim_{h \to 0-} \dfrac{s(t+h)-s(t)}{t+h-t}$ and $\lim_{h \to 0+} \dfrac{s(t+h)-s(t)}{t+h-t}$ must be isolated points for $s(t)$ to be traceable at $t$. If such points become dense in an infinitesimal neighborhood of $t$ then $s(t)$ will not be traceable in the whole neighborhood, as any attempt to trace $s(t)$ (say, by the marker of paper-ECG) at $t$ will jitter in random directions and will not move forward or backward. This will become clear in the following arguments, where we justify having only at most finitely many points in $(a,b)$, in which $s''(t)$ does not exist.

For $0 \le a < t < b < \infty$, let $s''(t)$ does not exist for an infinite number of points $t \in (a,b)$. Then $s''(t)$ does not exist for an infinite number of points $t \in [a,b]$. But by Heine-Borel theorem $[a,b]$ is a compact subset of the metric space $\Re^+ \cup \{0\}$

[32] (p. 114). Implies $[a,b]$ is sequentially compact (p. 124 of [32]), i.e., every set of infinitely many points has a convergent

subsequence. In other words, if the set $C = \{t_n \mid s''(t_n)\ \textit{does not exist}\}_{n=1}^{\infty}$ then it must have a subsequence that converges to a point $t' \in [a,b]$, such that, $s''(t')$ does not exist. We claim that there must be an open neighborhood $N_\delta(t')$, $\delta > 0$, so that

$s''(t)$ does not exist for any $t \in N_\delta(t')$.

Let the contrary be true, i.e., $\exists t_1 \in N_\delta(t')$ such that $s''(t_1)$ exists or $\lim_{h \to 0} \dfrac{s'(t_1+h)-s'(t_1)}{h}$ exists. Now, if we take

$\delta = h/2$ then $t' \in N_{h/2}(t_1)$. Let us choose $\alpha \in (0,1)$ such that $t_1 + \alpha h = t'$. So, $\lim_{h \to 0} s''(t'-\alpha h)$ exists or $s''(t')$ exists, leading to a contradiction.

When $s''(t)$ does not exist in an entire open neighborhood or an entire open interval of $\Re^+ \cup \{0\}$, even if $s'(t)$ is continuous in that interval, it cannot be traced, i.e., the two points $(t, s'(t))$ and $(t+h, s'(t+h))$, $h \to 0$ on the plane cannot be joined by a straight line each of whose points is of the form $(t', s'(t'))$. Let the contrary be true. Then $s'(t')$ will have a slope and $s''(t')$ will exist. This creates a situation in which the analog signal $s(t)$ becomes untraceable in that interval as elaborated below.

When $s''(t)$ does not exist in an open neighborhood $N_\delta(t)$, for an infinitesimal $\delta$, either $s'(t)$ does not exist or $s'(t)$ is non-differentiable in that neighborhood. In the first case $s(t)$ is not traceable in $N_\delta(t)$. This means, no plot of the signal $s(t)$ can be generated the way paper EEG or paper ECG is printed in $N_\delta(t)$. Also in $N_\delta(t)$ the signal cannot be modeled as the trajectory of a moving particle, violating a basic assumption about $s(t)$.

Let us consider the case when $s'(t)$ exists but nowhere differentiable in $N_\delta(t)$. By $s'(t)$ we get the gradient of the tangent to $s(t)$ at $t$, i.e., $s'(t) = \tan(\theta(t))$. $s''(t) = \sec^2(\theta(t))\theta'(t)$. $s''(t)$ does not exist anywhere in $N_\delta(t)$ means $\theta'(t)$ does not exist anywhere in $N_\delta(t)$. By the above argument $\theta(t)$ is not traceable in $N_\delta(t)$. But if $s(t)$ is to be traced from $t$ to $t+h$ as $h \to 0$, $\theta(t)$ will also have to be traced form $t$ to $t+k$ as $k \to 0$, or $\theta'(t)$ must exist, leading to a contradiction. This shows that if $s''(t)$ does not exist anywhere in $N_\delta(t)$, $s(t)$ is not traceable in $N_\delta(t)$ and $s(t)$ cannot be modeled as the trajectory of a moving particle in $N_\delta(t)$.

ACKNOWLEDGMENT

We would like to thank Anagh Pathak and Viswadeep Sarangi for some insightful computer simulations and logical arguments in the beginning of this work. This work was partly supported by an Indian Statistical Institute grant no. SSIU-03O-2014-2016 and a Department of Biotechnology, Government of India grant no. BT/PR7666/MED/30/936/2013.